\documentclass[twocolumn]{aastex631}

\shorttitle{Adiabatic compression of DM halos}
\shortauthors{Li et al.}

\graphicspath{{./}{figures/}}

\begin{document}

\title{The Effect of Adiabatic Compression on Dark Matter Halos and the Radial Acceleration Relation}

\correspondingauthor{Pengfei Li}
\email{PengfeiLi0606@gmail.com; pxl283@case.edu}

\author[0000-0002-6707-2581]{Pengfei Li}
\affiliation{Department of Astronomy, Case Western Reserve University, 10900 Euclid Avenue, Cleveland, OH 44106, USA}

\author[0000-0002-9762-0980]{Stacy S. McGaugh}
\affiliation{Department of Astronomy, Case Western Reserve University, 10900 Euclid Avenue, Cleveland, OH 44106, USA}

\author[0000-0002-9024-9883]{Federico Lelli}
\affiliation{INAF – Arcetri Astrophysical Observatory, Largo Enrico Fermi 5, I-50125, Firenze, Italy}

\author[0000-0001-9962-1816]{Yong Tian}
\affiliation{Institute of Astronomy, National Central University, Taoyuan 32001,
Taiwan}

\author[0000-0003-2022-1911]{James M. Schombert}
\affiliation{Department of Physics, University of Oregon, Eugene, OR 97403, USA}

\author[0000-0002-6459-4763]{Chung-Ming Ko}
\affiliation{Institute of Astronomy, National Central University, Taoyuan 32001,
Taiwan}
\affiliation{Department of Physics and Center for Complex Systems, National
Central University, Taoyuan 32001, Taiwan}

\begin{abstract}

We use a semi-empirical model to investigate the radial acceleration relation (RAR) in a cold dark matter (CDM) framework. Specifically, we build 80 model galaxies covering the same parameter space as the observed galaxies in the SPARC database, assigning them to dark matter halos using abundance matching and halo mass-concentration relations. We consider several abundance matching relations, finding some to be a better match to the kinematic data than others. We compute the unavoidable gravitational interactions between baryons and their dark matter halos, leading to an overall compression of the original NFW halos. Before halo compression, high-mass galaxies approximately lie on the observed RAR whereas low-mass galaxies display up-bending ``hooks" at small radii due to DM cusps, making them deviate systematically from the observed relation. After halo compression, the initial NFW halos become more concentrated at small radii, making larger contributions to rotation curves. This increases the total accelerations, moving all model galaxies away from the observed relation. These systematic deviations suggest that the CDM model with abundance matching alone cannot explain the observed RAR. Further effects (e.g., feedback) would need to counteract the compression with precisely the right amount of halo expansion, even in high mass galaxies with deep potential wells where such effects are generally predicted to be negligible.
\end{abstract}


\section{Introduction} \label{sec:intro}

Recent observations have revealed tight correlations between visible baryons and observed dynamics of galaxies, even in regions that are supposedly dominated by the dark matter (DM) halo \citep[e.g.,][]{McGaugh2020IAUS}. In the outermost galaxy regions, the circular speed along the flat part of the rotation curve correlate with galaxy luminosity, i.e., the Tully-Fisher relation \citep{TullyFisher1977}. The correlation becomes even tighter once galaxy luminosity is replaced by baryonic mass (stars plus gas), leading to the baryonic Tully-Fisher relation \citep[BTFR; e.g.,][]{McGaugh2000, Lelli2016}. In the innermost galaxy regions, the dynamical surface density given by the inner rise of the rotation curve correlates with the baryonic surface density, leading to a central density relation \citep[CDR;][]{Lelli2016c}. Moreover, at each galactic radius, the observed acceleration along the radial direction (${\rm g_{obs}}$) correlates with the gravitational field from the distribution of baryons (${\rm g_{bar}}$), leading to the radial acceleration relation \citep[RAR, e.g.][]{McGaugh2016PRL, OneLaw}. In particular, the RAR shows a characteristic acceleration scale ${\rm g_\dagger}$ below which the DM effect kicks in (${\rm g_{obs}>g_{bar}}$).

These findings question the DM paradigm as they suggest that baryonic matter ``knows'' exactly the whole kinematic behaviour of disk galaxies. It raises a conspiracy problem in the DM context: DM halos and baryonic disks have to closely collaborate (conspire) to make those correlations appear. On the other hand, these empirical relations were predicted a-priori in the context of Milgromian Dynamics \citep[MOND;][]{Milgrom1983}, in which the classical laws of Newtonian dynamics are modified at low accelerations instead of adding DM.

Several groups have tried to reproduce the observed RAR in a CDM cosmology using hydrodynamic simulations of galaxy formation \citep{Keller2017, Ludlow2017, Tenneti2018, Garaldi2018, Dutton2019}. All these studies find a correlation between $\mathrm{g}_{\rm tot} = \mathrm{g}_{\rm bar} + \mathrm{g}_{\rm DM}$ and $\mathrm{g}_{\rm bar}$, which is mathematically expected because the two quantities are not independent in simulated galaxies (contrary to the observational situation). The resulting correlation, however, is not necessarily comparable to the observed RAR. For example, \citet{Tenneti2018} finds that the $\mathrm{g}_{\rm tot}$--$\mathrm{g}_{\rm bar}$ relation is linear with no sign of a characteristic acceleration scale, while \citet{Ludlow2017} report a value of ${\rm g_\dagger}=2.6\times10^{-10}$ m/s$^2$. This is significantly higher than the observed one \citep{Lelli2017} with a formal discrepancy of 70$\sigma$ (random) and 5.8$\sigma$ (systematic). \citet{Ludlow2017} address this issue by doubling their stellar mass, and hence $\mathrm{g}_{\mathrm{bar}}$, which translates their relation into approximate agreement with the data. However, we will show that there is not really freedom to do this because $\mathrm{g}_{\rm DM}$ is also impacted: it is not a simple translation along one axis. Furthermore, it is not trivial to compare the intrinsic scatter of the observed RAR with that from simulated galaxies \citep{Keller2017, Garaldi2018, Dutton2019} because one needs to model observational errors, rotation-curve sampling, and the co-variance between ${\rm g_{tot}}$ and ${\rm g_{bar}}$.

A different approach to study the RAR is using semi-empirical analytic models in which baryonic disks are assigned to DM halos using abundance-matching prescriptions \citep{DiCintio2016, Desmond2017, Navarro2017, Grudic2020, Paranjape2021}. To achieve a fair comparison, model galaxies are supposed to match the properties of observed ones, such as baryonic mass and size. In addition, the model rotation curves have to be cut at both large and small radii to match the ranges that observed rotation curves cover. For example, \citet{Navarro2017} removed the inner parts at $R<0.747R_{\rm d}$ ($R_{\rm d}$ is the disk scale length), which, however, are present in many observed rotation curves from the SPARC database \citep{SPARC}. In fact, the inner parts of rotation curves are of great interest because the classical cusp-vs-core problem appears at small radii \citep[e.g., see][]{deBlok2008, Adams2014, Oh2015, Katz2017}, so it is unclear how the models of \citet{Navarro2017} compare to observations.

When setting up baryonic disks and DM halo separately, one neglects the mutual gravitational interaction between these two components. The models of \citet{Navarro2017} completely neglect baryonic effects, assuming that DM halos preserve the ``original'' NFW profile from DM-only cosmological simulations \citep{Navarro1996}. In contrast, \citet{DiCintio2016} found a reasonable match to the RAR using a halo profile from hydrodynamic simulations \citep{DiCintio2014} that takes core formation from stellar feedback into account. Core formation is most efficient at stellar masses of $\sim$10$^{7-8}$ M$_\odot$, leaving cusps in the halos of more massive galaxies with stellar masses of $\sim$10$^{10-11}$ M$_\odot$ \citep[][their fig. B1]{Katz2017}. \citet{Desmond2017} took the approach of using a free parameter to account for a global halo contraction or expansion, finding it difficult to explain the detailed shape and small scatter of the RAR.

We wish to rigorously compute the gravitational interaction between baryons and DM halos as galaxies form: one should not simply drop a fully formed galaxy into a static DM halo. This is a necessary step to insure dynamical stability. \citet{Sellwood2005} examined this problem by investigating the response of DM halos to the growth of baryonic disks. They numerically computed the profiles of the stabilized DM halos, finding that initial NFW halos experienced adiabatic contraction due to baryonic compression \citep[see also][]{Abadi2010}. This adiabatic compression leads to larger DM contributions to the final rotation curves and can alter the shape of the RAR \citep[see][]{Paranjape2021}. \citet{Dutton2007} found that adiabatically contracted models cannot simultaneously reproduce the observed Tully-Fisher relation, size-luminosity relation, and luminosity function.

In this paper, we investigate the RAR in a CDM cosmology using the semi-empirical approach, including a proper treatment of the compression of DM halos. We pose a simple and intentionally limited question: what happens if we compute the expected compression for models matched to the data? We do not attempt to model further effects (e.g., feedback), nor claim to explain the RAR. Rather, we attempt to establish the prior expectation for a natural model prior to the complicating effects of baryonic physics \citep[e.g.,][]{Chan2015} in the hope of elucidating the specific outcomes that these need to accomplish.

The paper is organized as follows: Section 2 describes our approach to build model galaxies; Section 3 presents their rotation curves and the corresponding RAR in the absence of adiabatic compression; Section 4 describes the algorithm for implementing baryonic compression and illustrates how this affects the model rotation curves and the resultant RAR. We discuss our results in Section 5 and provide a brief summary in Section 6. 

\section{Model setup}
\begin{figure*}
    \centering
    \includegraphics[scale=0.41]{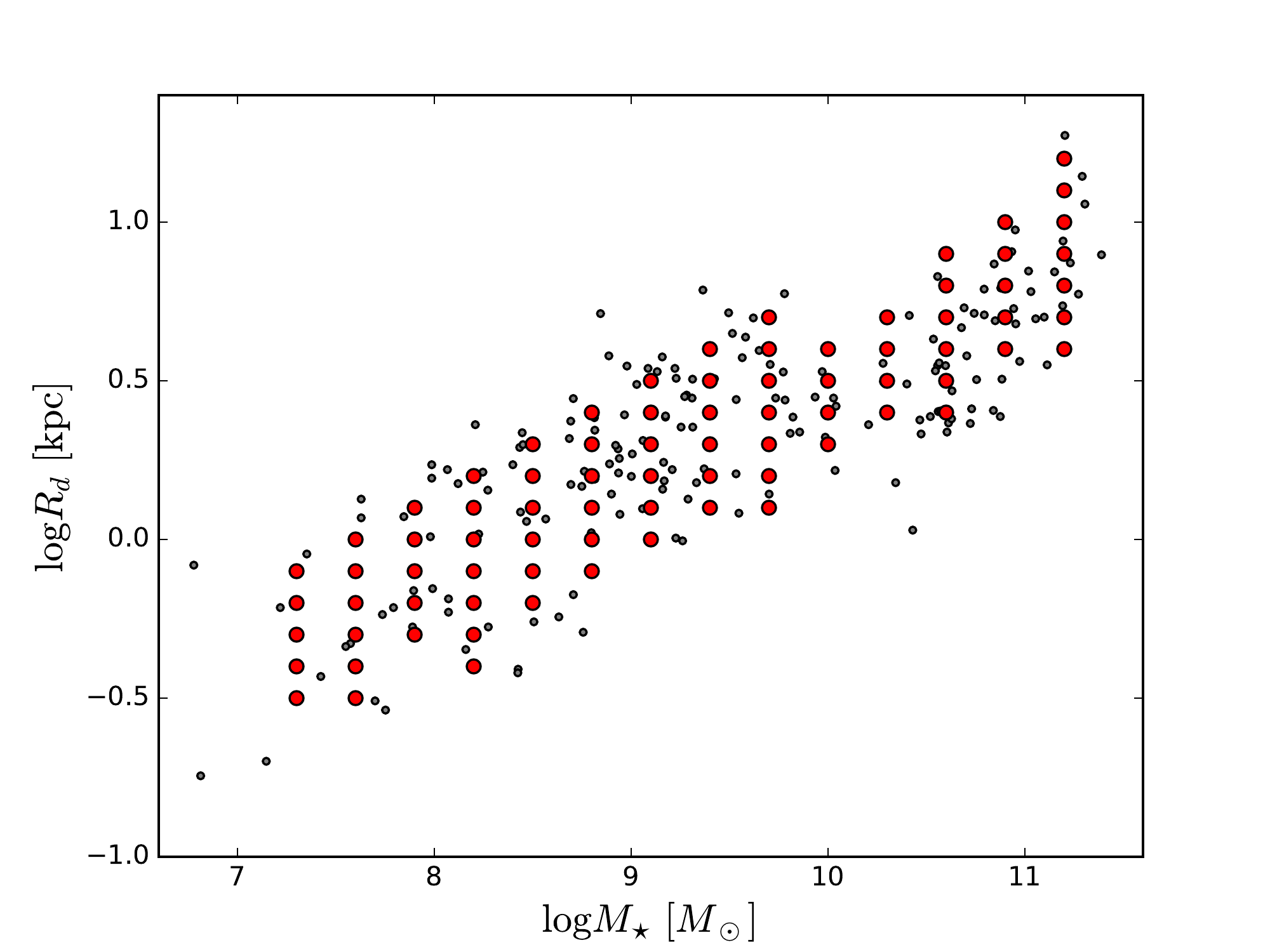}\includegraphics[scale=0.41]{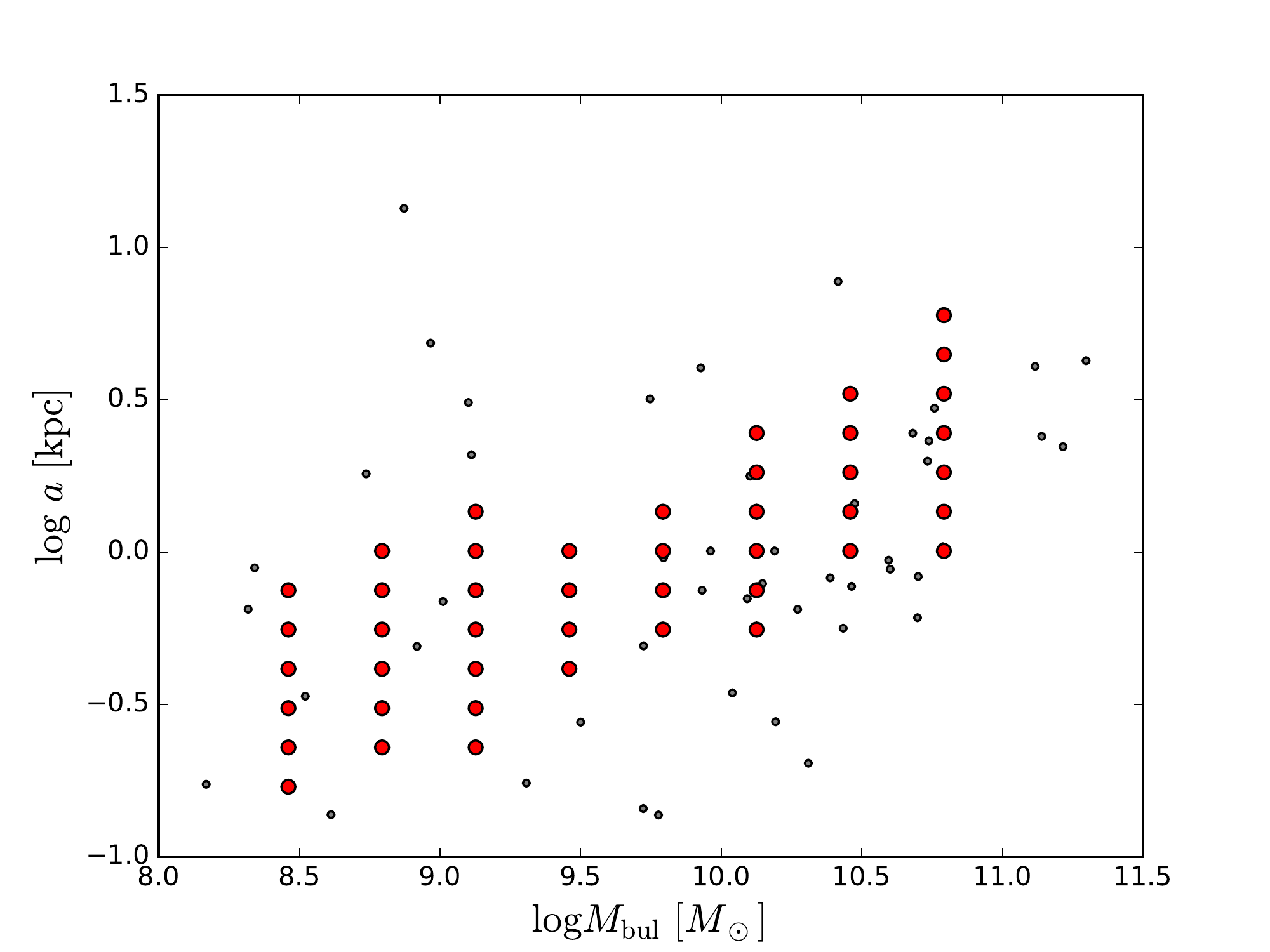}
    \caption{The parameter space of model galaxies (red points) compared with real galaxies (black points) from the SPARC database. In the left panel, the total stellar mass $M_\star$ and disk scale length $R_{\rm d}$ of model galaxies is chosen to match those of real galaxies (assuming a stellar mass-to-light ratio of 0.5 at 3.6 $\mu$m). In the right panel, the bulge mass $M_{\rm bul}$ and bulge scale length $a$ of model galaxies with $M_\star\geq10^9\ M_\odot$ are derived using empirical mass-mass and size-size relations (see the text for details).}
    \label{fig:ParameterSpace}
\end{figure*}

\subsection{Stellar disks}

Late-type galaxies typically present a thin stellar disk, whose surface mass density can be approximated with an exponential function:
\begin{equation}
    \Sigma(R) = \Sigma_0e^{-R/R_{\rm d}},
\end{equation}
where $\Sigma_0$ is the central surface density and $R_{\rm d}$ is the disk scale length. The total stellar mass is therefore
\begin{equation}
    M_{\rm d} = 2\pi\Sigma_0R^2_{\rm d}.
\end{equation}
The gravitational contribution to the rotation curve is given by \citep{Freeman1970}:
\begin{equation}
    V^2_{\rm d} = \frac{GM_{\rm d}}{R_{\rm d}}2y^2[I_0(y)K_0(y) - I_1(y)K_1(y)],
\end{equation}
where $I_n$ and $K_n$ are the modified Bessel functions of the first and second kind, respectively, and $y=R/(2R_{\rm d})$. The surface mass density profiles and rotation curves are fully determined by two parameters: total stellar mass $M_{\rm d}$ and disk scale length $R_{\rm d}$.

The left panel of Figure \ref{fig:ParameterSpace} shows the $M_{\star}$-$R_{\rm d}$ parameter space covered by the SPARC sample, assuming a stellar mass-to-light ratio of 0.5 at 3.6 $\mu$m. To make our sample comparable to the SPARC data, we model galaxies covering the same parameter space and build a sample that includes 80 galaxies, equally space in both $\log M_\star$ and $\log R_{\rm d}$. For galaxies with a bulge, their disk masses are calculated by subtracting the bulge masses from the total stellar masses. 

\subsection{Central bulges}

Bulges are a common feature of early-type spiral galaxies that are often neglected in models based on exponential disks. Bulges are important because they dominate the high acceleration ($> 10^{-9} \;\mathrm{m}\,\mathrm{s}^{-2}$) regime of the RAR where $\mathrm{g}_{\mathrm{tot}} \simeq \mathrm{g}_{\mathrm{bar}}$. A model lacking bulges cannot explain this regime.

For models of massive galaxies ($M_{\star}\geq10^9\ M_\odot$), we build two samples, one with a central bulge and one without. The central bulge is built using the Hernquist profile \citep{Hernquist1990}:
\begin{equation}
    \rho(r) = \frac{M_{\rm bul}a}{2\pi r(r+a)^3},
\end{equation}
where $M_{\rm bul}$ is the total bulge mass and $a$ is the characteristic radius within which the enclosed bulge mass equals to $M_{\rm bul}/4$. We determine the bulge mass using the empirical correlation between bulge and total stellar masses from SPARC galaxies:
\begin{equation}
    \log M_{\rm bul} = 1.11\log M_\star - 1.64.
\end{equation}
To determine the size parameter $a$, we use the disk size-bulge size relation,
\begin{equation}
    \log a = 1.29\log(R_{\rm d}) - 0.77,
\end{equation}
which is derived from fitting the disk and bulge sizes of the SPARC galaxies. This is not a strong correlation, but it provides a realistic range of $a$ values (Figure \ref{fig:ParameterSpace}).

Bulges populate the high acceleration end of the RAR; they are star dominated with no clear need for dark matter. As such, a different choice of bulge model (equation 4), or variations on equations 5 and 6, or the scatter in size at a given mass, only occur where $\mathrm{g}_{\mathrm{tot}} \simeq \mathrm{g}_{\mathrm{bar}}$ for plausible models. Consequently, our results do not strongly depend on the precise form of these relations.

The cumulative mass distribution of the Hernquist profile is given by 
\begin{equation}
    M(r) = \frac{M_{\rm bul}r^2}{(r+a)^2},
\end{equation}
and its contribution to rotation velocities is
\begin{equation}
    V_{\rm bul} = \sqrt{\frac{GM_{\rm bul}r}{(r+a)^2}}.
\end{equation}
This suffices to calculate the compression of a model since we assume that the bulge component is spherical. However, for consistency with the treatment of real data for which only the projected light distribution is observed, we compute the projected surface brightness of the Henquist bulge and use this as the input to \texttt{compress}. Integrating the density along the line of sight, one obtains \citep[e.g.,][]{Hernquist1990}:
\begin{equation}
    \Sigma_{\rm bul}(R) = \frac{M_{\rm bul}}{2\pi R_{1/4}^2(1-s^2)^2}\big[(2+s^2)X(s)-3\big],
\end{equation}
where $s=R/a$ with $R$ being the projected radius, and 
\begin{equation}
    X(s) = \frac{\ln[(1+\sqrt{1-s^2})/s]}{\sqrt{1-s^2}}
\end{equation}
for $0 \leq s\leq 1$;
\begin{equation}
    X(s) = \frac{\arccos(1/s)}{\sqrt{s^2-1}}
\end{equation}
for $1\leq s \leq \infty$. At $s=1$, $X(s=1)=1$, the surface mass density has a finite value
\begin{equation}
    \Sigma_{\rm bul}(R=a) = \frac{2M_{\rm bul}}{15\pi a^2}.
\end{equation}

\subsection{Gas disks}

Assuming a razor-thin exponential disk, we model the gas contribution with two parameters: total gas mass $M_{\rm gas}$ and gas scale length $R_{\rm gas}$. The total gas mass is derived through the gas mass-stellar mass relation as in \citet{Chae2021}:
\begin{equation}
    M_{\rm gas} = X^{-1} (11500M_\star^{0.54} + 0.07 M_\star)
\end{equation}
with masses in solar units. This includes both atomic and molecular hydrogen gas \citep{SPARC}. To account for helium, we multiply the derived mass by the factor of $1.33$. Following \citet{DiCintio2016}, we specify the gas scale length using the empirical relation $R_{\rm gas} = 2R_{\rm d}$. The gas contribution is important in dwarf galaxies and in the outermost parts of some spiral galaxies. In both cases, g$_{\rm tot}$ is largely dominated by the DM halo, so the precise modeling of the gas distribution does not strongly affect our results, having only minor effects on g$_{\rm bar}$ in the low-acceleration regime.

\subsection{The Circumgalactic Medium}

In addition to cold gas, galaxies are surrounded by coronae of warm-hot gas. This circumgalactic medium (CGM) may contain a baryonic mass comparable to that in the stars \citep{Werk2014,Tumlinson2017,Bregman2021} albeit in a diffuse form distributed over a much larger volume (100s of kpc). We do not attempt to model this component as it is not detected in individual SPARC galaxies, so the model RAR would not be compatible with the observed RAR. Indeed, the extended, diffuse nature of the CGM ensures that it will contribute negligibly to both axes of the RAR over the tens of kpc typically spanned by the SPARC data. There is some hint from lensing data \citep{Brouwer2021} that the CGM might become important at larger scales and lower accelerations than where we are interested here.

\subsection{Dark matter halos}

DM halos from N-body cosmological simulations can be described by the NFW halo \citep{Navarro1996}:
\begin{equation}
    \rho(r) = \frac{\rho_s}{\frac{r}{r_s}(1+\frac{r}{r_s})^2},
\end{equation}
where $\rho_s$ is the characteristic volume density and $r_s$ is the scale radius. It is common to describe NFW halos using two parameters defined at the so-called ``virial radius''
$r_{200}$ (within which the average halo mass density is 200 times the critical density of the Universe): the halo concentration $C_{200}$ and the halo mass $M_{200}$. These two parameters are defined as
\begin{equation}
    C_{200} = r_{200}/r_s; \ M_{200} =100\ r^3_{200}H^2_0/G, 
\end{equation}
where $G$ and $H_0 = 73$ km s$^{-1}$ Mpc$^{-1}$ are Newton and Hubble constants, respectively. We will use these two parameters to describe NFW halos in this paper, and determine their values using $M_\star$-$M_{200}$ relations (constant mass ratio or abundance matching) and halo mass-concentration relations.

\begin{figure*}
    \centering
    \includegraphics[scale=0.41]{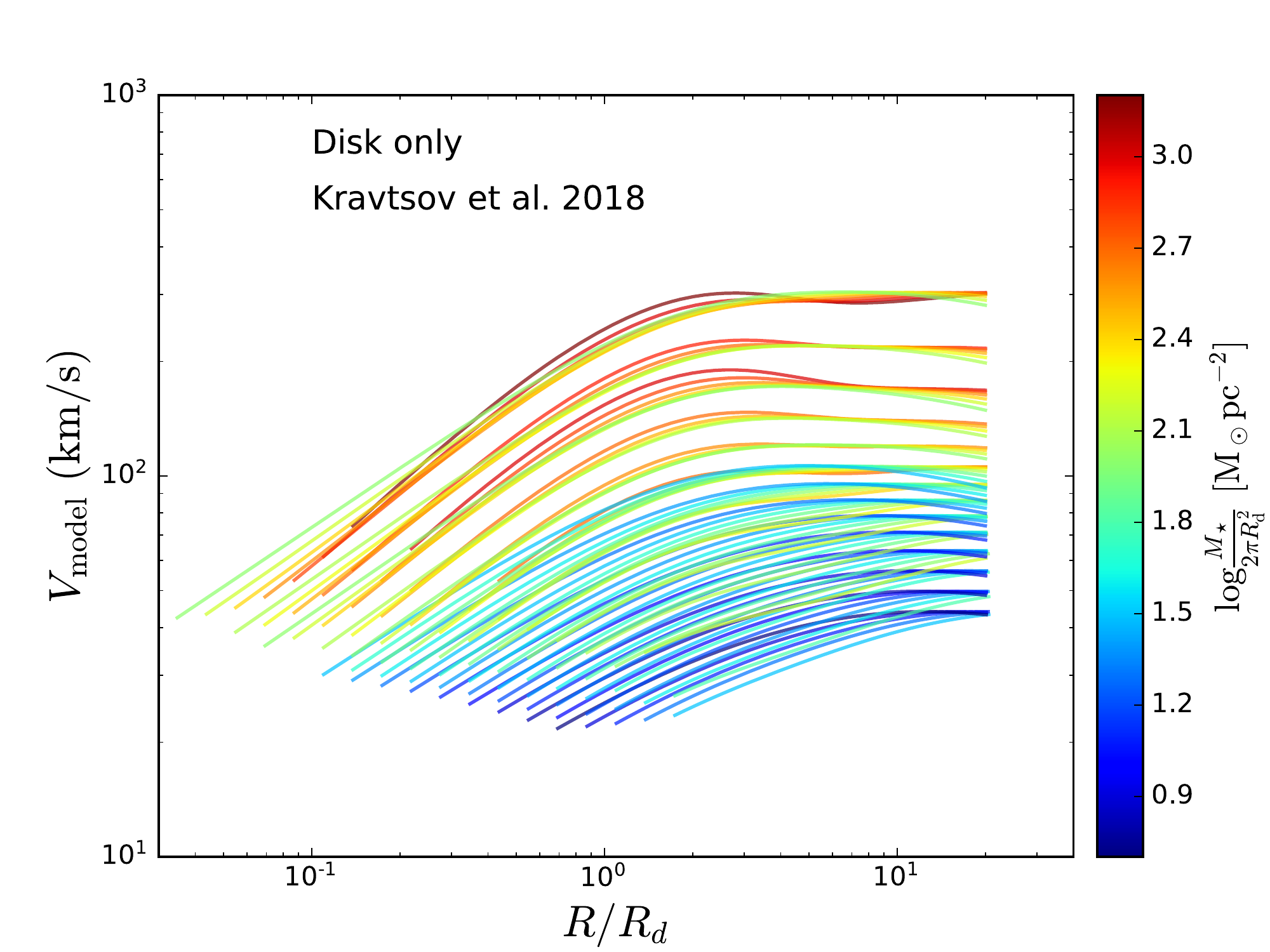}\includegraphics[scale=0.41]{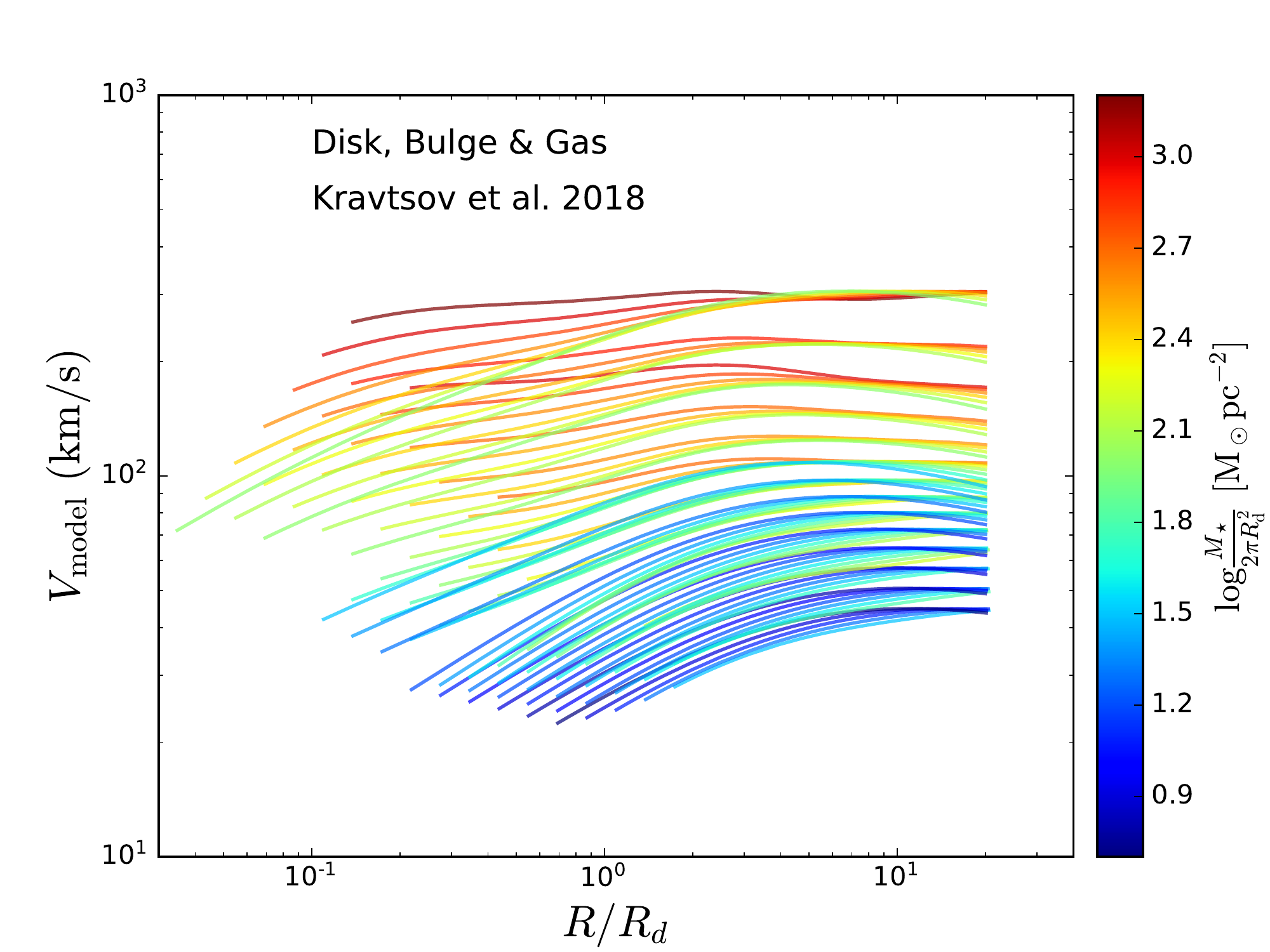}
    \includegraphics[scale=0.41]{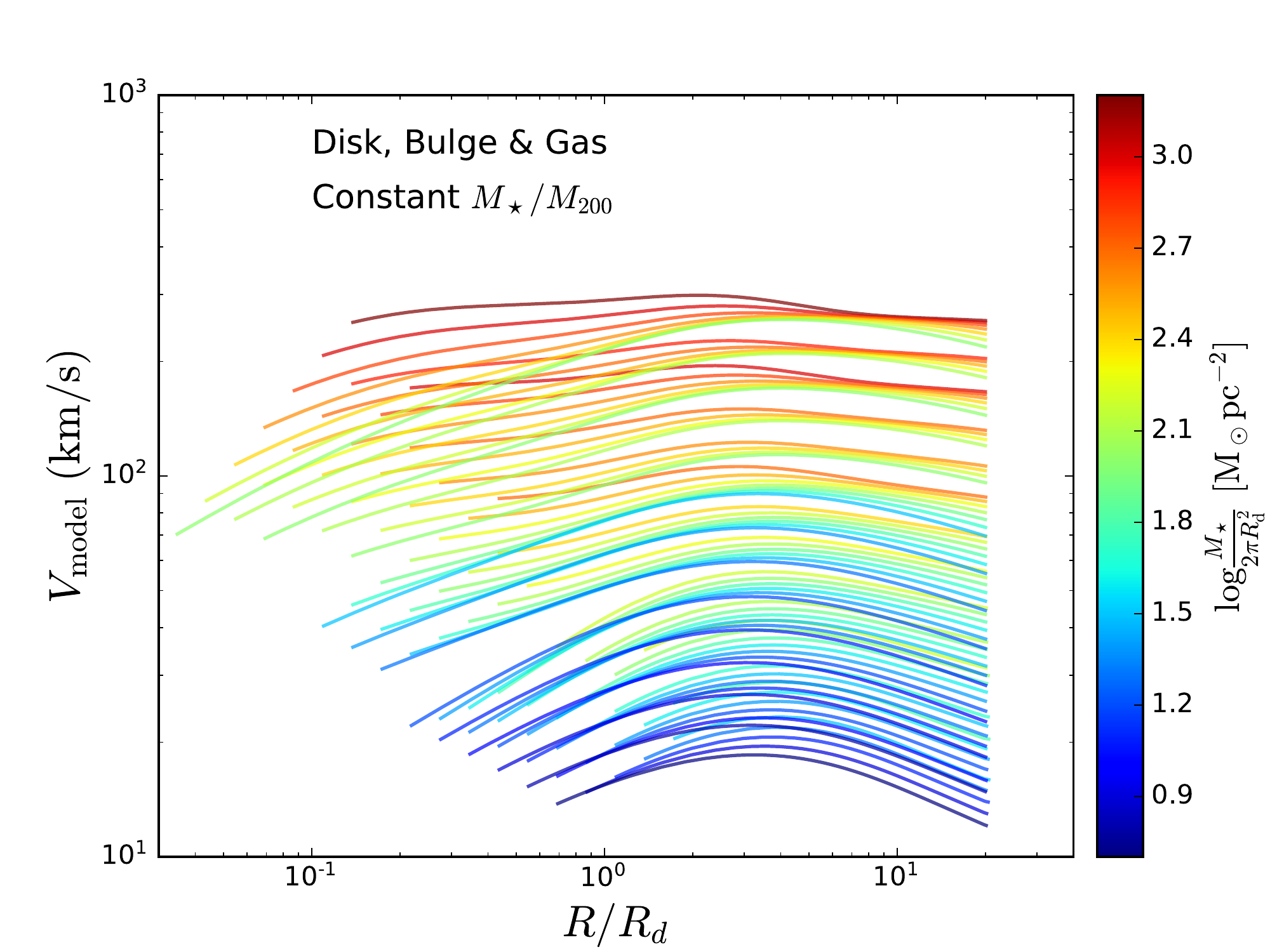}\includegraphics[scale=0.41]{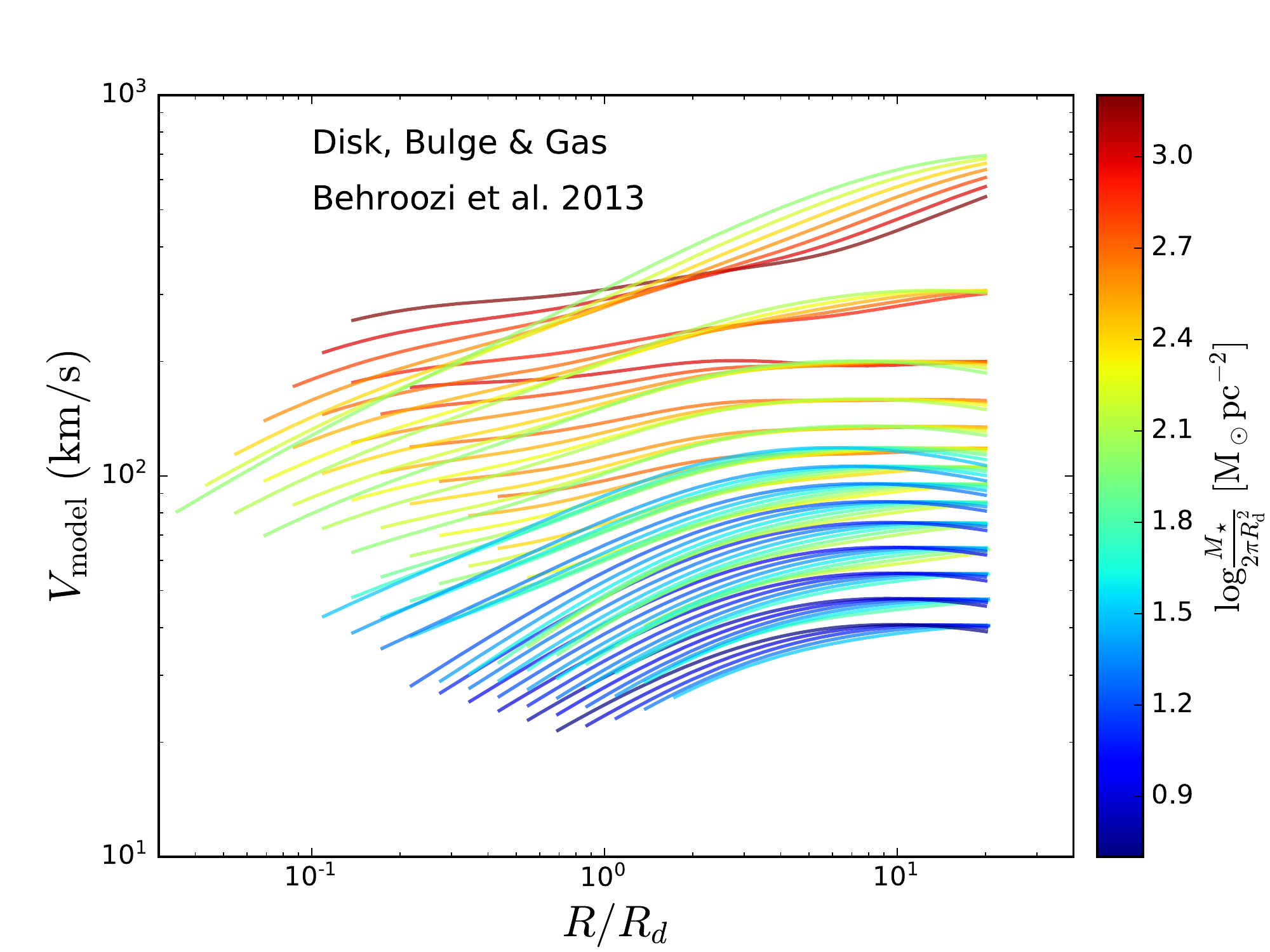}
    \caption{The rotation curves of the model galaxies assuming different $M_\star$-$M_{200}$ relations. Galaxies are color coded by $\frac{M_\star}{2\pi R_{\rm d}^2}$ (i.e. central surface stellar-mass densities for disk-only galaxies). The model rotation curves are cut at 0.5 kpc in the inner parts \citep[see Figure 3 in][]{McGaugh2020IAUS} and 20 $R_{\rm d}$ in the outer parts, according to the range the SPARC rotation curves cover.}
    \label{fig:RotVel}
\end{figure*}

\begin{figure*}
    \centering
    \includegraphics[scale=0.41]{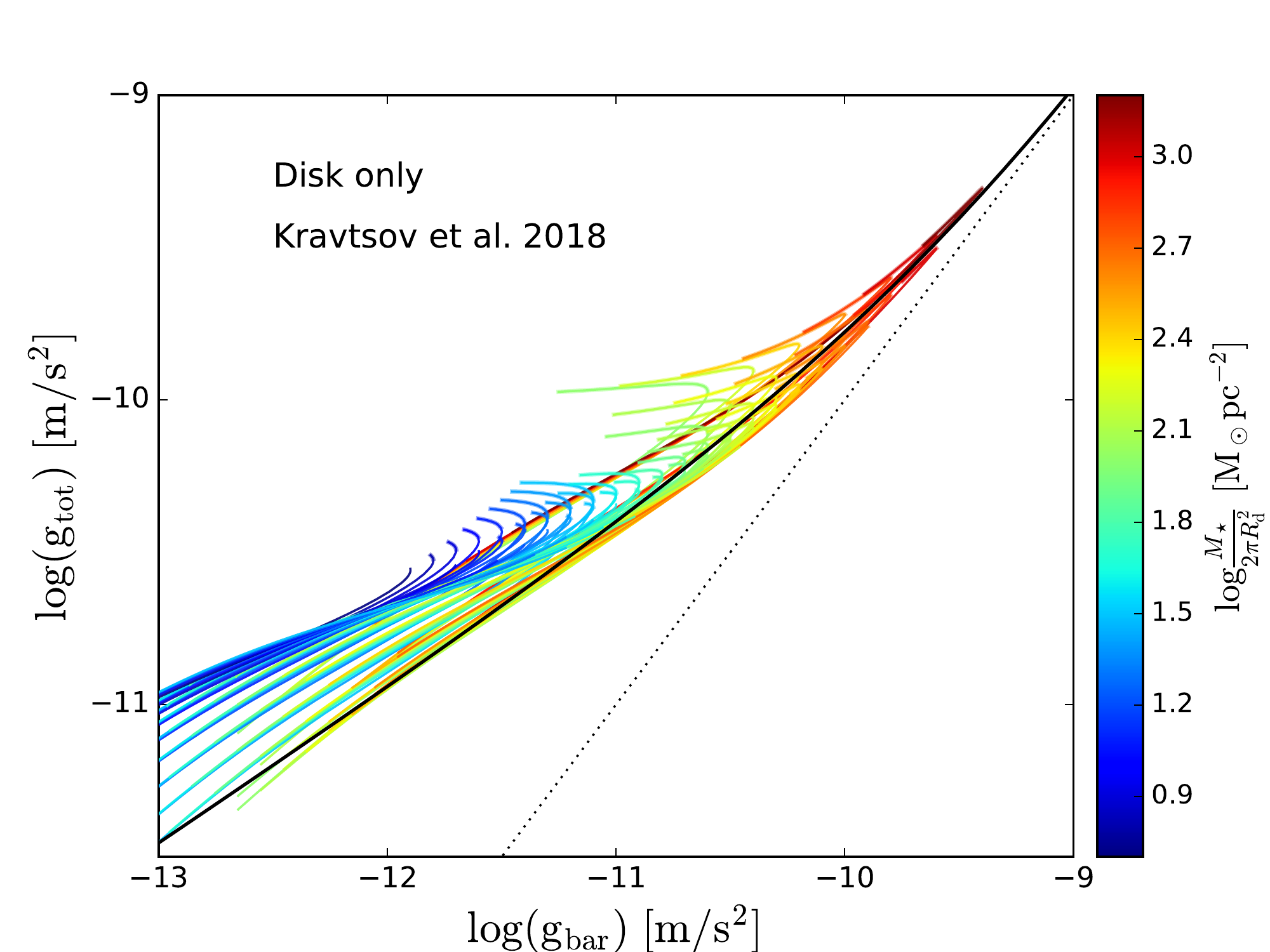}\includegraphics[scale=0.41]{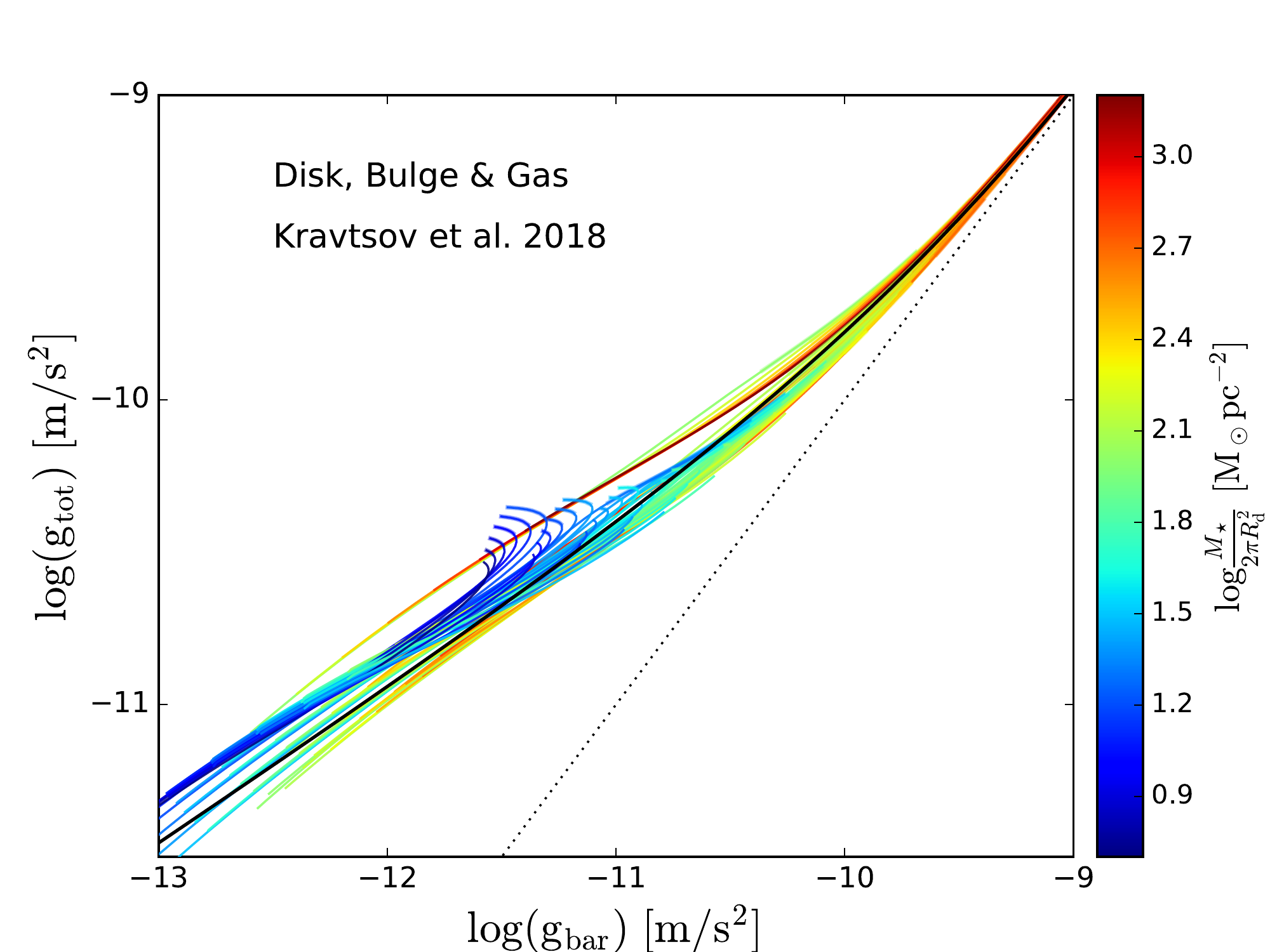}
    \includegraphics[scale=0.41]{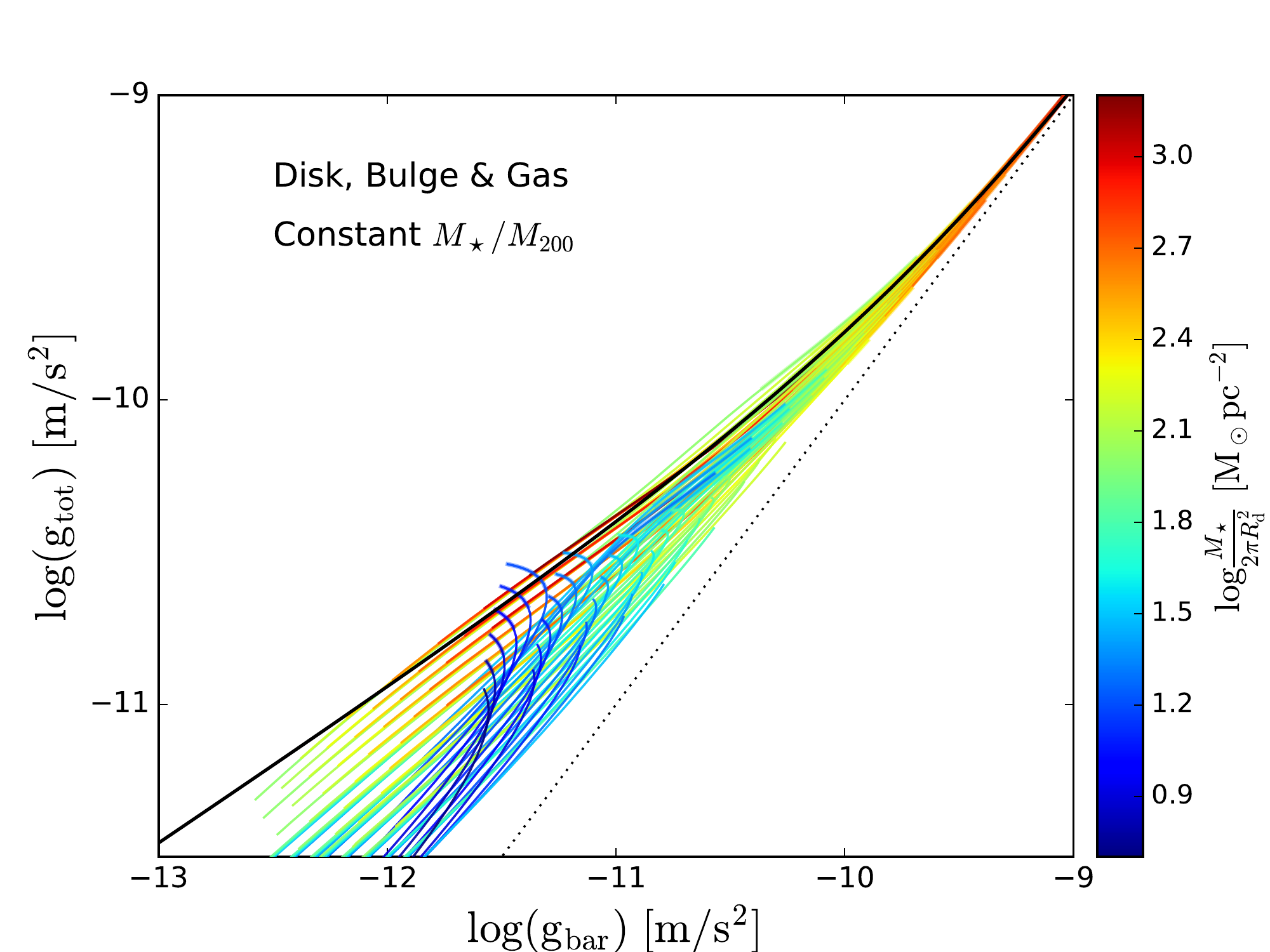}\includegraphics[scale=0.41]{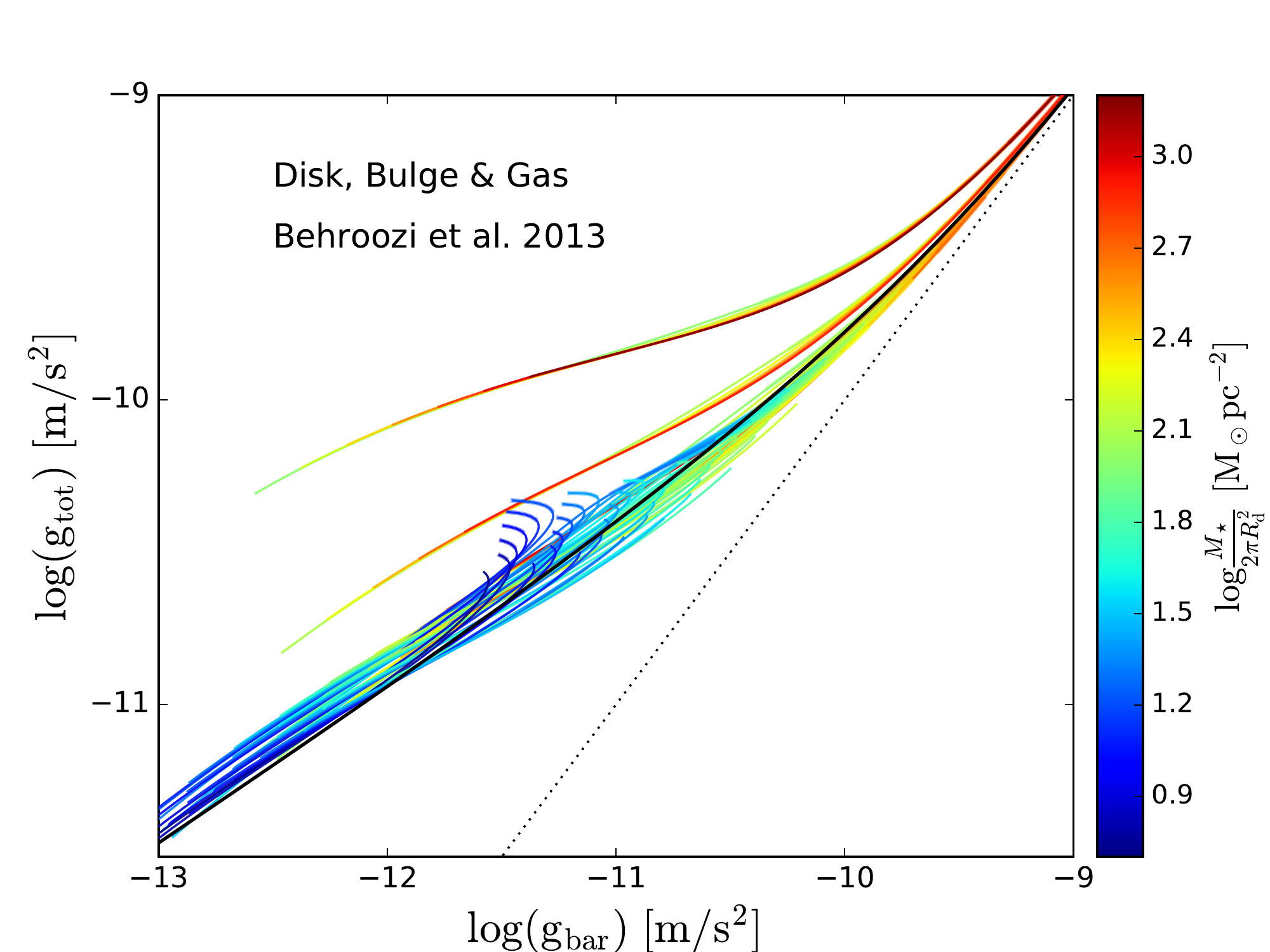}
    \caption{The predicted RAR using different $M_\star$-$M_{200}$ relations. Black solid lines represent the observed RAR \citep{McGaugh2016PRL, Lelli2016}, and dotted lines are the line of unity. The disk-only model predicts a hook shaped RAR for individual galaxies with a turning point at $R=0.747R_d$. The inner parts at $R<0.747R_{\rm d}$ with increased opacity and thickness bend upwards showing larger discrepancy from the observed RAR. The central bulges in massive galaxies effectively move the turning point further into the center and out of the observable range. The predicted RAR for low-mass galaxies is either systematically lower (constant $M_\star/M_{200}$) or higher (abundance matching relations) than the mean observed relation.}
    \label{fig:RAR_NFW}
\end{figure*}

\subsubsection{Setting the DM Halo Mass: A Constant Mass Ratio}

The total mass of a dark matter halo remains a difficult quantity to ascertain. A classic approach \citep[e.g.,][]{DSS97} is to assume a constant ratio of normal to dark matter. Here we adopt
\begin{equation}
    M_\star = 0.05M_{200}.
\end{equation}
consistent with the findings of \citet{Mo1998}.
Models of this type fail badly \citep{McGaughdeBlok1998}, causing far too much scatter in the RAR. We include such a model here as an important historical point of reference.

\subsubsection{Setting the DM Halo Mass: Abundance Matching}

It is now widely recognized that a constant baryon fraction is not viable in terms of either kinematics \citep{McGaugh2010,Posti2019b} or the number density of galaxies \citep{Guo2010,Moster2013,Behroozi2013}. In the latter case, a constant stellar mass to halo mass ratio cannot reproduce the observed stellar mass function of galaxies (a Schechter function), starting from the theoretical DM halo mass function. Reproducing the observed stellar mass function requires a non-linear correlation between stellar and halo mass that has become known as abundance matching.

Abundance matching has now become a popular tool to determine the stellar mass-halo mass relation (SMHMR). Observational methods using satellite kinematics \citep[e.g.][]{Conroy2007} and weak lensing \citep[e.g.][]{Mandelbaum2006, Leauthaud2012, Hudson2015} have also been used to measure the SMHMR. Here, we consider different SMHMRs from abundance matching and apply them to kinematic data.

One curious aspect of abundance matching relations is the nonlinearity of the relation between stellar and halo masses above and below the break in the Schechter function. Below the break, a large range of stellar mass is compressed into a narrow range of halo mass. Above the break, the situation is reversed, and a little stellar mass range goes a long way in terms of halo mass. This can be problematic, as there is a proclivity for abundance matching to predict DM halo masses for bright galaxies that are too large for the observed kinematics \citep[e.g.,][]{McGaughvanDokkum2021}. In contrast, we expect the compression of the range of halo masses for low mass galaxies to be helpful insofar as it is a step in the direction of making the DM halos of rather different galaxies look similar, as indicated by kinematics \citep{McGaugh2007}. Whether this is merely a numerical convenience, or how the universe really works is a larger question.

Various abundance matching relations have been proposed. These have non-negligible differences at both the high-mass and low-mass ends \citep[e.g.,][]{Guo2010, Moster2013, Moster2018, Kravtsov2021}. In this work, we consider two abundance matching relations that brackets most of the relations proposed in the literature: \citet{Behroozi2013} and \citet{Kravtsov2018}. Both studies adopted the same parameterization function,
\begin{equation}
\log M_\star = \log(\epsilon M_1) + f\Big[\log\Big(\frac{M_{200}}{M_1}\Big)\Big] - f(0),
\end{equation}
where the function $f(x)$ is defined as 
\begin{equation}
    f(x) = -\log(10^{\alpha x} + 1) + \delta\frac{[\log(1+e^x)]^\gamma}{1+e^{10^{-x}}}.
\end{equation}
\citet{Behroozi2013} found that $\alpha=-1.412$, $\gamma=0.316$, $\delta=3.508$, $\log M_1=11.514$, and $\log\epsilon=-1.777$. This gives a relation close to that from \citet{Moster2013}. Instead, \citet{Kravtsov2018} found a significantly different behaviour at the high-mass end with the following parameters: $\alpha=-1.779$, $\gamma=0.547$, $\delta=4.394$, $\log M_1 = 11.35$, $\log\epsilon=-1.642$. \citet{DiCintio2016} found that the \citet{Kravtsov2018} relation describes high-mass galaxies better than \citet{Moster2013} or \citet{Behroozi2013} because a galaxy with $M_\star\simeq10^{11}$ M$_\odot$ is assigned to a DM halo with $M_{200}\simeq10^{13}$ M$_\odot$ rather than $\sim$10$^{14-15}$ M$_\odot$. As such, we use the \citet{Kravtsov2018} relation as our fiducial relation in this paper, and present the results from other $M_\star$-$M_{200}$ relations as reference. 

\subsubsection{Setting the DM Halo Concentration}

To derive halo concentrations, we use the halo mass-concentration relation, which has been extensively studied in galaxy formation simulations. This relation depends on cosmology and is sensitive to halo profiles \citep{Maccio2008, DuttonMaccio2014}. In the WMAP5 cosmology, the halo mass-concentration relation for the NFW profile is given by
\begin{equation}
    \log(C_{200}) = 0.830 - 0.098\log(M_{200}/[10^{12}h^{-1}M_\odot]).
\end{equation}
Both the abundance-matching relation and halo mass-concentration relation have significant scatter, but we assume accurate match when sampling. With these two relations, we completely determine the DM halo for a given stellar mass. 

\section{The RAR for non-compressed halos}

The total rotation curves of model galaxies are given by the quadratic sum of the various mass contributions:
\begin{equation}
    V^2_{\rm model} = V^2_{\rm DM} + V^2_{\rm d} + V^2_{\rm bulge}.
\end{equation}
Figure \ref{fig:RotVel} plots the rotation curves for four cases: disk-only galaxies with the abundance matching relation (AMR) from \citet{Kravtsov2018}, disk$+$bulge$+$gas galaxies with the AMR from \citet{Kravtsov2018}, disk$+$bulge$+$gas galaxies with a constant $M_\star/M_{200} = 0.05$ from \citet{Mo1998}, and disk$+$bulge$+$gas galaxies with the AMR from \citet{Behroozi2013}. For illustration, we truncate the model rotation curves at $20R_{\rm d}$, comparable to the most extended rotation curves in the SPARC database. At small radii, the SPARC rotation curves are mostly observed at $r>$ 0.5 kpc, so we apply this as the inner cutoff. Throughout the paper, we will use the same cutoff for all rotation curves and the RAR.

We color code rotation curves according to $\log\frac{M_\star}{2\pi R_{\rm d}^2}$, which is the central surface stellar-mass density for disk-only galaxies. As expected, galaxies with higher central surface mass densities rotate faster. In the disk-only case, the rotation curves are roughly self-similar because of the self-similarity of exponential disks and NFW halos, but high-mass galaxies rise too slowly in the inner parts with respect to real galaxies. When a bulge component is added for massive galaxies, the inner parts of the rotation curves become flat and more comparable to the SPARC rotation curves. This shows that the central bulge is an essential component to model a massive galaxy.

The imposed $M_\star$-$M_{200}$ relations have a significant effect on the rotation curves. For low-mass galaxies, assuming $M_\star/M_{200}=0.05$ leads to systematically smaller rotation velocities than abundance matching. This occurs because abundance-matching relations imply lower stellar fractions ($M_\star/M_{200}\ll0.05$) for dwarf galaxies, so the assumption $M_{200}=M_\star/0.05$ assign a dwarf galaxy with given $M_\star$ to a less massive DM halo. For high-mass galaxies, the \citet{Kravtsov2018} relation and the constant $M_\star/M_{200}$ model give similarly flat rotation curves. The \citet{Behroozi2013} relation, however, presents much larger rotation velocities at large radii for high-mass galaxies because it predicts larger halo masses for a given stellar mass at high-mass end. This would imply rising rotation curves at large radii, which are not observed in real galaxies.

We calculate ${\rm g_{tot}}$ and ${\rm g_{bar}}$ from the model rotation curves and the baryonic contributions, respectively. They are plotted for each individual galaxy in Figure \ref{fig:RAR_NFW}. Disk-only galaxies show a ``hook'' in the RAR that consistently bends upwards due to an excess of DM at small radii. A similar phenomenon has been observed in \citet{Lelli2017} when testing the emergent gravity theory of \citet{Verlinde2011}. In our CDM models, this occurs because exponential disks have a maximum value of ${\rm g_{bar}}$ at $R=0.747R_{\rm d}$. At smaller radii, the value of ${\rm g_{bar}}$ decreases towards the center. Given that the NFW model is cuspy, DM halos make significantly larger contributions to rotation curves at smaller radii. This leads to considerably larger total acceleration ${\rm g_{tot}}$, bending the curves upwards. As such, the upwardly bending hook is a reflection of the core-cusp problem. A central bulge can significantly increase the baryonic mass density at small radii, so it effectively moves the radius at which the baryonic acceleration peaks further in. How far the turning point moves depends on how compact and massive the bulge is. All the modeled bulges effectively move their turning points out of the observable range. As a result, massive galaxies with a bulge present a non-hooked RAR.

The RAR of low-mass galaxies is more sensitive to $M_\star$-$M_{200}$ relations because they are DM dominated. The constant $M_\star/M_{200}$ model shows insufficient gravitational contributions from DM halos. One can adopt a lower $M_\star/M_{200}$ ratio to achieve a better match but the systematic hooks would persist. The \citet{Behroozi2013} relation implies very high halo mass for the most massive galaxies in our sample, so the total accelerations are larger than expected at all radii. In fact there are some models that lie well above the observed RAR. The DM halos derived using the \citet{Kravtsov2018} relation can better reproduce the observed RAR at the high-mass end.

\begin{figure}
    \centering
    \includegraphics[scale=0.41]{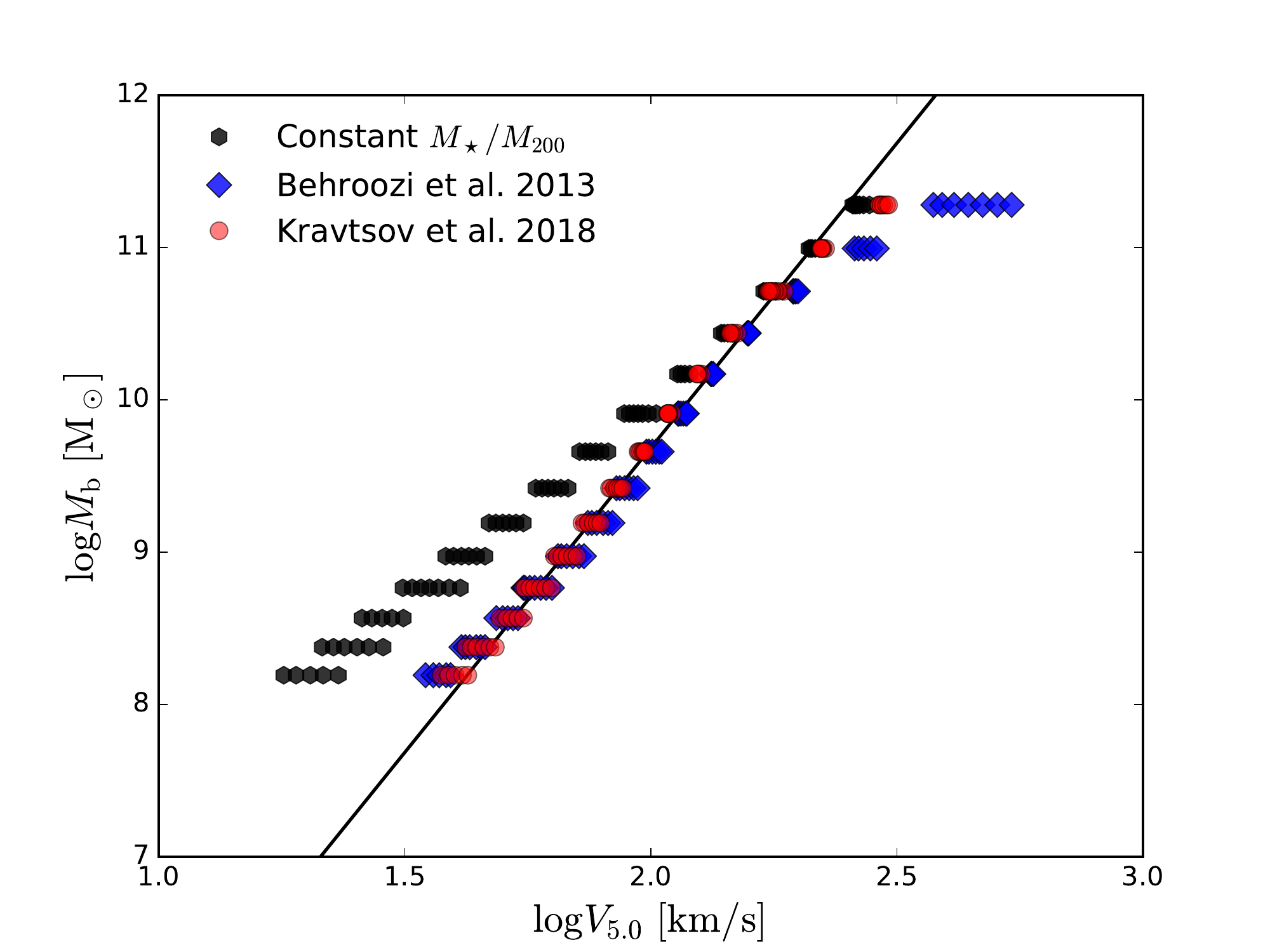}
    \caption{The baryonic Tully-Fisher relations (BTFR) from different $M_\star$-$M_{200}$ relations. The solid line is the calibrated BTFR \citep{Schombert2020} with a slope of 4 \citep{McGaugh2021}. The constant $M_\star/M_{200}$ model recovers a slope of $\sim3.0$ as expected from the halo mass--velocity relation \citep{SteinmetzNavarro}. Models employing the abundance matching relations provide a reasonable match to the observed BTFR below $L^*$. Above this scale, the BTFR for bright galaxies is predicted to bend to higher velocities, consistent with the bend in the $M_\star-M_{200}$ relations. This bend is not observed \citep{superspirals}.}
    \label{fig:BTFR}
\end{figure}

To further investigate the effect of the $M_\star$-$M_{200}$ relation, we check their corresponding baryonic Tully-Fisher relations (BTFR) in Figure \ref{fig:BTFR}. The solid line is the BTFR calibrated by \citet{Schombert2020},
\begin{equation}
M_b = (48.5\;\mathrm{M}_{\odot}\,\mathrm{km}^{-4}\,\mathrm{s}^4)\; V_f^4. \label{eq:BTFR}
\end{equation}
Since the model rotation curves may not remain flat at large radii, there is not a clear definition for $V_{f}$. For illustrative purpose, we choose the velocities at $R=5R_{\rm d}$, given most of the SPARC rotation curves have become flat by this radius. We check that using a different radius does not affect our conclusion.

At large radii, DM halos dominate the rotation curves, so $M_{200}\sim V^3$. This implies that the imposed $M_\star$-$M_{200}$ relation will imprint its shape onto the BTFR plot. When we assume a constant $M_\star/M_{200}$, it leads to a linear BTFR but with a slope of $\sim3$ as expected. Abundance matching, instead, predicts a bend in the BTFR. The \citet{Kravtsov2018} relation is less bent, and so also is the resultant BTFR. The \citet{Behroozi2013} relation bends significantly at the high mass end, so we observe a big increase in the rotation velocities. A bent abundance matching relation is therefore in conflict with the linear baryonic Tully-Fisher relation. 
\begin{figure}
    \centering
    \includegraphics[scale=0.41]{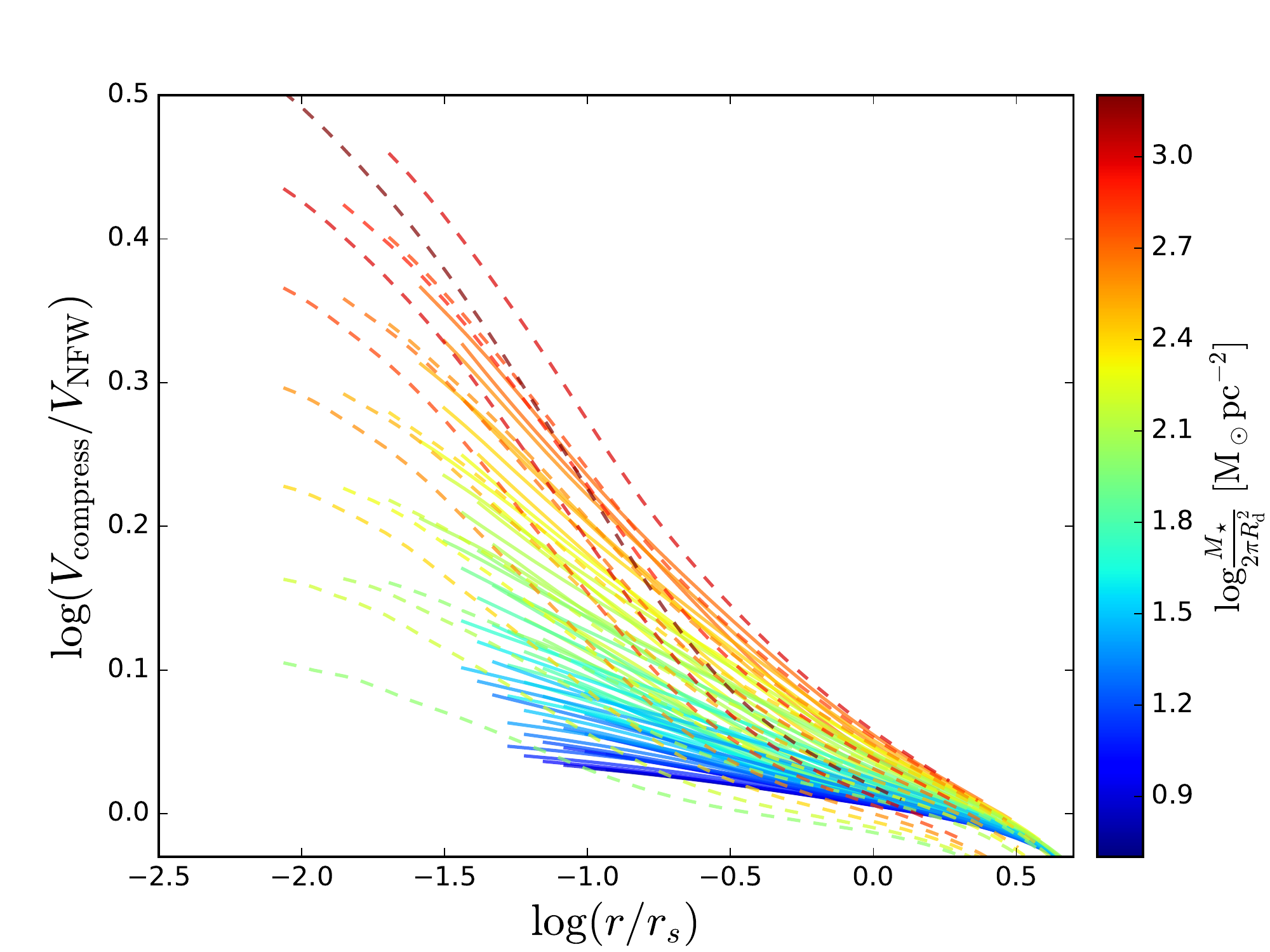}
    \caption{The ratios of the rotation velocities of compressed halos to pure NFW halos using the abundance matching relation from \citet{Kravtsov2018}. Solid/dashed lines represent galaxies with stellar mass smaller/larger than $10^{10.5}$ $M_\odot$, color-coded by $\frac{M_\star}{2\pi R_{\rm d}^2}$. Solid lines show a perfect color gradient: higher surface mass densities present stronger compression effects. The break at $10^{10.5}$ $M_\odot$ originates from the turning point in the abundance match relation.}
    \label{fig:Vratio}
\end{figure}

\section{Baryonic compression of DM halos}

In the previous Sections, we treated the baryonic components and DM halos separately, ignoring their mutual interaction. In this Section, we use the \texttt{compress} code \citep{Sellwood2014} to compute the response to the primordial NFW halo to the formation of the baryonic galaxy. The final halo is no longer exactly NFW in form, but the combined system is in dynamical equilibrium. 
\subsection{The compression code}

The DM halo must respond to the changing gravitational potential that results from the growth of the baryonic galaxy at its center. This evolution is often modeled as adiabatic, which is a good approximation even in the hierarchical cosmogony of $\Lambda$CDM \citep{Choi2006}. \citet{Young1980} introduced an algorithm that conserves all three adiabatic actions ($J_r$, $J_\phi$, $J_\theta$). This is a major improvement over previous studies \citep[e.g.][]{Barnes1984, Blumenthal1986, Ryden1987}, which overstate the effects of compression \citep{Sellwood1999, Gnedin2004}. Conserving the third action, the radial adiabatic invariant, effectively makes DM halos less susceptible to compression.

\begin{figure*}
    \centering
    \includegraphics[scale=0.41]{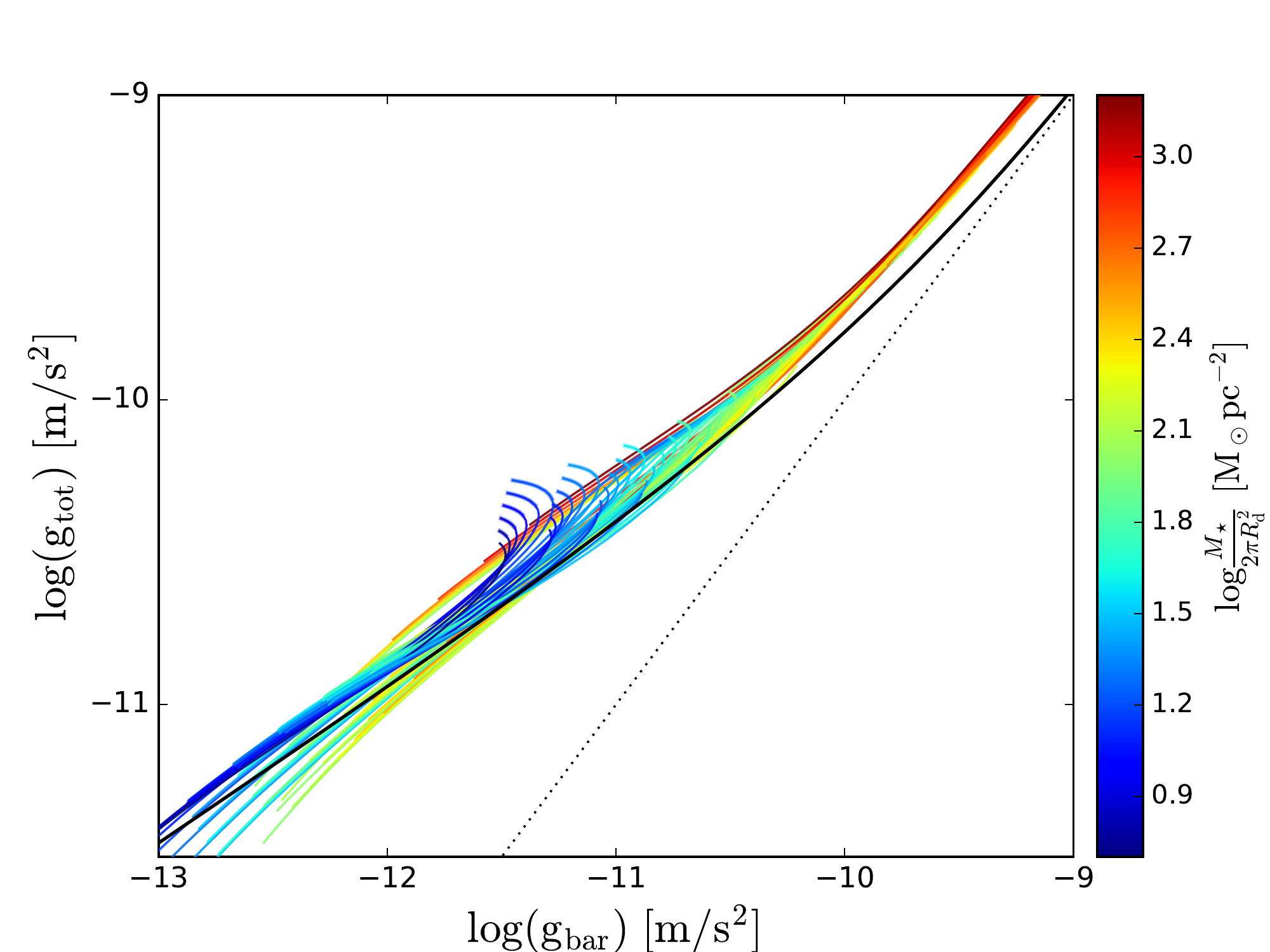}\includegraphics[scale=0.41]{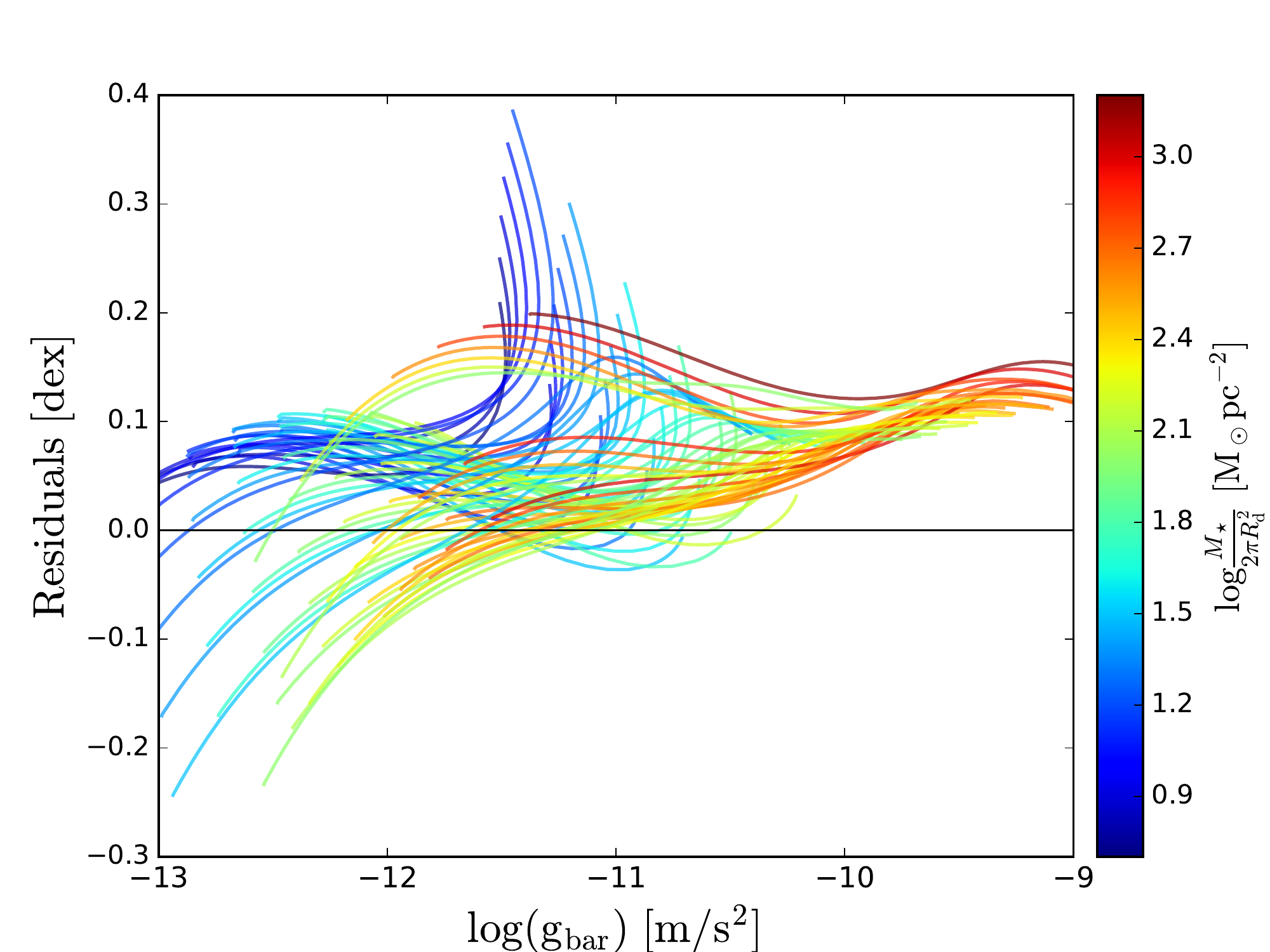}
    \caption{The RAR (left) and its residuals (right) when baryonic compression is considered assuming the abundance matching relation from \citet{Kravtsov2018}. Symbols are the same as that in Figure \ref{fig:RAR_NFW}. The model shows a systematic deviation above the RAR. Compression effects are more significant at small radii, so that massive galaxies present a big upwards shift with respect to Figure \ref{fig:RAR_NFW}.}
    \label{fig:RAR_compress}
\end{figure*}

\citet{Sellwood2005} developed Young's algorithm and applied it to the adiabatic contraction of DM halos. It starts with a given NFW halo and updates its density profile iteratively as baryons are added. Throughout the process, spherical symmetry is assumed, which significantly simplifies the computation. Its validity has been tested with N-body simulations by \citet{Jesseit2002}. Using \texttt{compress}, we calculate the compressed DM halos for all 80 model galaxies that are built from the \citet{Kravtsov2018} relation. 

\subsection{Rotation curves and RAR from compressed halos}

To illustrate the effects of baryonic compression, Figure \ref{fig:Vratio} shows the ratios of the rotation velocities of the compressed halos to that of the original NFW halos. The degree of halo compression depends strongly on the baryonic surface mass density. The halos of low-surface-density galaxies experience almost negligible adiabatic contraction, while those of high-surface-density galaxies are strongly compressed. As expected, baryonic compression is more significant at smaller radii, where the surface mass density is higher. 

Galaxies with $m_\star>10^{10.5}$ $M_\odot$ (dashed lines in Fig.\,\ref{fig:Vratio}) respond rather differently to baryonic compression. This occurs because they lie in a different regime of the abundance-matching relation, in which DM halo mass increases more rapidly with stellar mass. The resulting stellar-to-halo mass ratio becomes smaller, so that baryonic effects turn to be less sufficient to compress massive halos than that expected in the low-mass region.

The baryonic effects have a direct implication for the RAR because larger rotation velocities correspond to larger total accelerations. The RAR from compressed halos and its residuals are shown in Figure \ref{fig:RAR_compress}. As mentioned earlier, the halos of low-surface-density galaxies experience less significant adiabatic contraction, so their location on the RAR remains roughly the same (see Figure \ref{fig:RAR_NFW}). The halos of high-surface-density galaxies suffer strong baryonic compression, and the compression is more profound at small radii due to the strong gravitational pull of the central bulges. The net result is that their total accelerations are increased considerably after halo contraction, presenting a large discrepancy from the observed RAR.

The residual plot better illustrates the discrepancy. All model galaxies are clearly above the baseline across almost all radii. Since the NFW model predicts a falling rotation curve at large radii, some galaxies fall below the baseline. The ``hooks'' in low-mass galaxies becomes more distinct, given that the compressed halos are more concentrated at small radii. High-mass galaxies, though not hooked, experience substantial baryonic contraction, and so never approach the one-to-one line observed at high accelerations.

\section{Discussion}
\label{sec:disc}

We have made a careful calculation of adiabatic compression for model galaxies well matched to the observed range of size and mass present in the SPARC data that define the RAR. A model with a constant stellar-to-halo mass ratio \citep[as in][]{Mo1998} produces a RAR with an enormous amount of scatter, much greater than observed. This is, in part, why such models were rejected as unworkable by \citet{McGaughdeBlok1998}. More recent models built on the same principle \citep[e.g.,][]{DiCintio2016, Desmond2017, Navarro2017} appear more successful because the assumption of a constant stellar-to-halo mass ratio has been replaced with a variable ratio from abundance matching. This has the effect of reducing the range of halo masses for the range of observed stellar masses, reducing the scatter in the RAR \citep{DiCintio2016}.

Models that we built with abundance matching relations have lower scatter in the RAR for the same reason. However, the scatter is not entirely negligible, and none of the models provide an entirely satisfactory match to the observed RAR. The problems encountered differ for high and low mass galaxy models.

Abundance matching relations have a bend around a halo mass of $10^{12}\;\mathrm{M}_{\odot}$. This imprints a scale on the predicted kinematics that is not clearly observed. In the Tully-Fisher plane, abundance matching models perform well for low mass galaxies, but over-predict the velocities of high mass galaxies. The severity of this problem depends on how sharp the bend in the abundance matching relation is. Of the models considered here, those built with the relation of \citet{Kravtsov2018} suffer less from this effect than those built with the relation of \citet{Behroozi2013}. This problem manifests in the RAR plane as models with excessively large total accelerations.

A further problem that arises at high accelerations is that models with adiabatically compressed halos never reach the one-to-one line where $\mathrm{g}_{\mathrm{bar}} = \mathrm{g}_{\mathrm{tot}}$. This problem does not appear if we ignore compression; a sufficiently high surface brightness bulge or disk will be sufficiently dominant so that $\mathrm{g}_{\mathrm{bar}} \approx \mathrm{g}_{\mathrm{tot}}$ at $\mathrm{g}_{\mathrm{bar}} \gg \mathrm{g}_{\dagger}$ and the model looks fine \citep[e.g.,][]{Navarro2017}. Unfortunately, this is an artifact of ignoring the inevitable gravitational response of the dark matter halo to the formation of the luminous galaxy. If we start from the initial NFW halos predicted by DM-only simulations, then the high concentrations of baryons required to reach high $\mathrm{g}_{\mathrm{bar}}$ have a strong effect on the central cusp of the dark matter halo, which compresses to maintain $\mathrm{g}_{\mathrm{tot}} > \mathrm{g}_{\mathrm{bar}}$. Consequently, one never expects to reach the observed one-to-one line.

Low-mass galaxy models form a sequence that might be considered a reasonable approximation of the observed RAR. However, these models generically display ``hooks'' at small radii. These are the manifestation of the cusp-core problem in bulgeless galaxy models. The centripetal acceleration $\mathrm{g}_{\mathrm{bar}}$ of an exponential disk has a maximum at a radius of $0.747 R_{\rm d}$. This marks a maximum along the abscissa of the RAR plane for pure disk models. In contrast, the acceleration contributed by the DM halo increases monotonically to the center. Following any given model from outside in, one follows a track of increasing acceleration in the RAR plane until the maximum in $\mathrm{g}_{\mathrm{bar}}$ is reached. The models then bend back, hooking away from the observed RAR as $\mathrm{g}_{\mathrm{tot}}$ continues to increase. These features\footnote{Hooks are are not readily apparent in the models of \citet{Navarro2017} because they discontinued their plots at $0.747 R_{\rm d}$.} are a generic prediction of disk models in NFW halos.

The data do not clearly display the predicted upwardly hooking behavior. For example, the residuals around the mean RAR do not correlate with radius or with $R/R_{\rm eff}$, where $R_{\rm eff}$ is the stellar half-mass radius \citep{Lelli2017}. There are many of examples of galaxies in SPARC that have well resolved rotation curves within $0.747 R_{\rm d}$, so this behavior should be apparent if present. The predicted effect is not subtle.

Great care must be taken in evaluating both $\mathrm{g}_{\mathrm{bar}}$ and $\mathrm{g}_{\mathrm{tot}}$ at small radii. The first is sensitive to the adopted stellar mass-to-light ratio of the stars while the second is sometimes subject to resolution effects. Indeed, there are some hints of hooks in the data, but at a much smaller amplitude than predicted. We do not believe these features to be meaningfully significant, for the reasons just given. They scatter both upwards and downwards\footnote{While we find that pure CDM models predict upturned hooks due to the central cusp of the NFW profile, Bullock (2017, private communication) noted that SIDM predicts downturned hooks.}, as expected for noise.

\section{Conclusion}
\label{sec:conc}

In this paper, we examined the RAR expected in $\Lambda$CDM by building 80 model galaxies. The models include both stars and gas, with exponential disks and Hernquist bulges chosen to match the observed range of properties in the SPARC database \citep{Lelli2016}. Model galaxies are embedded in NFW halos. We consider several prescriptions for relating the DM halo mass to the stellar mass, and calculate the effect of baryonic compression to elucidate the expected effects of the formation of the luminous galaxy on the mass distribution of the DM halo.

We identify at least three distinct problems: (i) the shape of the stellar mass-halo mass relation, which over-predicts velocities for high-mass galaxies; (ii) the inability of our adiabatically compressed models to reach the one-to-one line at high accelerations; (iii) the presence of upturned but unobserved hooks in the predicted RAR for low-mass galaxy models.

The first of these problems seems minor and yet intractable. It is minor insofar as the variation of the stellar mass--halo mass relation obtained from abundance matching is approximately correct to explain the same variation indicated by kinematics. It is intractable in that there is inevitably a bend in abundance matching relations that simply is not present in kinematic data. Abundance matching imprints a mass scale $M_{200} \approx 10^{12}\;\mathrm{M}_{\odot}$ while kinematics evince an acceleration scale $\mathrm{g}_{\dagger} \approx 10^{-10}\;\mathrm{m}\,\mathrm{s}^{-2}$. The two are not obviously related. The mass scale of abundance matching predicts a bend in the plane of the baryonic Tully-Fisher relation where the acceleration scale is defined by this relation as a single power law \citep[eq.\ \ref{eq:BTFR}; see][]{McGaugh2020IAUS}.

Problems (ii) and (iii) might conceivably be addressed with further baryonic physics that we have not considered, like feedback. It is an intentional choice on our part to separate these effects so that we can see what compression alone does. This sets the standard for what feedback (or other baryonic physics) needs to do in order to reconcile $\Lambda$CDM models with the observations. However, there are different problems that appear in distinct regimes, so it is not obvious that generic feedback effects will solve all problems. Indeed, many works that might resolve the cusp-core problem \citep[e.g.,][]{DiCintio2014} have their strongest effects on low to intermediate mass galaxies, while leaving high mass galaxies unscathed. This addresses (iii) while leaving (ii) unaddressed. In any case, whatever mechanism is invoked must precisely undo the effects of compression at high surface densities without inducing too much scatter in the RAR. This unavoidably requires fine tuning between gravitational and non-gravitational baryonic effects on DM halos. 

A deeper issue that we do not address is why the relations that we struggle to explain here, the BTFR and the RAR, were successfully predicted \textit{a priori} by MOND \citep{Milgrom1983}.

\begin{acknowledgments}
We are grateful to Jerry Sellwood for help and advice in using the \texttt{COMPRESS} code. The work of PL, SSM, JMS are supported in part by NASA ADAP grant 80NSSC19k0570. SSM also acknowledges support from NSF PHY-1911909. YT is supported by the Taiwan Ministry of Science and Technology grant MOST 110- 2112-M-008-015-MY3. CMK is supported by grant MOST 110-2112-M-008-005.

\end{acknowledgments}

\bibliography{PLi}{}
\bibliographystyle{aasjournal}

\end{document}